\documentclass[letterpaper, 10 pt, conference]{ieeeconf}  

\IEEEoverridecommandlockouts      

\usepackage{amsmath} 
\usepackage{graphicx}
\usepackage{booktabs}
\usepackage{amssymb}
\usepackage{siunitx}
\usepackage[english]{babel}
\usepackage[noadjust]{cite}
\usepackage{units}
\usepackage{epstopdf}
\usepackage[font=small]{caption}
\usepackage{subcaption}
\usepackage{comment}
\usepackage{amsfonts}
\usepackage{bm}
\usepackage{algorithm}

\usepackage{amsthm}
\usepackage{algorithmic}

\usepackage{color}

\usepackage{xcolor}
\newcommand{\remove}[1]{}

\theoremstyle{plain}

\newtheorem{theorem}{Theorem}

\newtheorem{definition}[theorem]{Definition}

\title{\LARGE \bf
Responsibility-associated Multi-agent Collision Avoidance \\
with Social Preferences
}

\author{Yiwei Lyu$^1$, Wenhao Luo$^2$ and John M. Dolan$^3$
\thanks{$^*$This work was supported in part by the CMU Argo AI Center for Autonomous Vehicle Research and the Faculty Research Grant award at UNC Charlotte.}
\thanks{$^1$The author is with the Department of Electrical and Computer Engineering, Carnegie Mellon University. Email: {\tt \small yiweilyu@andrew.cmu.edu}}%
\thanks{$^{2}$The author is with the Department of Computer Science, University of North Carolina at Charlotte. Email: {\tt \small wenhao.luo@uncc.edu}}%
\thanks{$^{3}$The author is with the Robotics Institute, Carnegie Mellon University. Email: {\tt \small jmd@cs.cmu.edu}}%
}

\begin{document}

\maketitle
\thispagestyle{empty}
\pagestyle{empty}

\begin{abstract}
 
This paper introduces a novel social preference-aware decentralized safe control framework to address the responsibility allocation problem in multi-agent collision avoidance. Considering that agents do not necessarily cooperate in symmetric ways, this paper focuses on semi-cooperative behavior among heterogeneous agents with varying cooperation levels. Drawing upon the idea of Social Value Orientation (SVO) for quantifying the individual selfishness, we propose a novel concept of Responsibility-associated Social Value Orientation (R-SVO) to express the intended relative social implications between pairwise agents. This is used to redefine each agent's social preferences or personalities in terms of corresponding responsibility shares in contributing to the coordination scenario, such as semi-cooperative collision avoidance where all agents interact in an asymmetric way. By incorporating such relative social implications through proposed Local Pairwise Responsibility Weights, we develop a Responsibility-associated Control Barrier Function-based safe control framework for individual agents, and multi-agent collision avoidance is achieved with formally provable safety guarantees. Simulations are provided to demonstrate the effectiveness and efficiency of the proposed framework in several multi-agent navigation tasks, such as a position-swapping game, a self-driving car highway ramp merging scenario, and a circular position swapping game.

\end{abstract}

\section{Introduction}

\begin{figure}
    \centering
    \includegraphics[width = 0.9\linewidth]{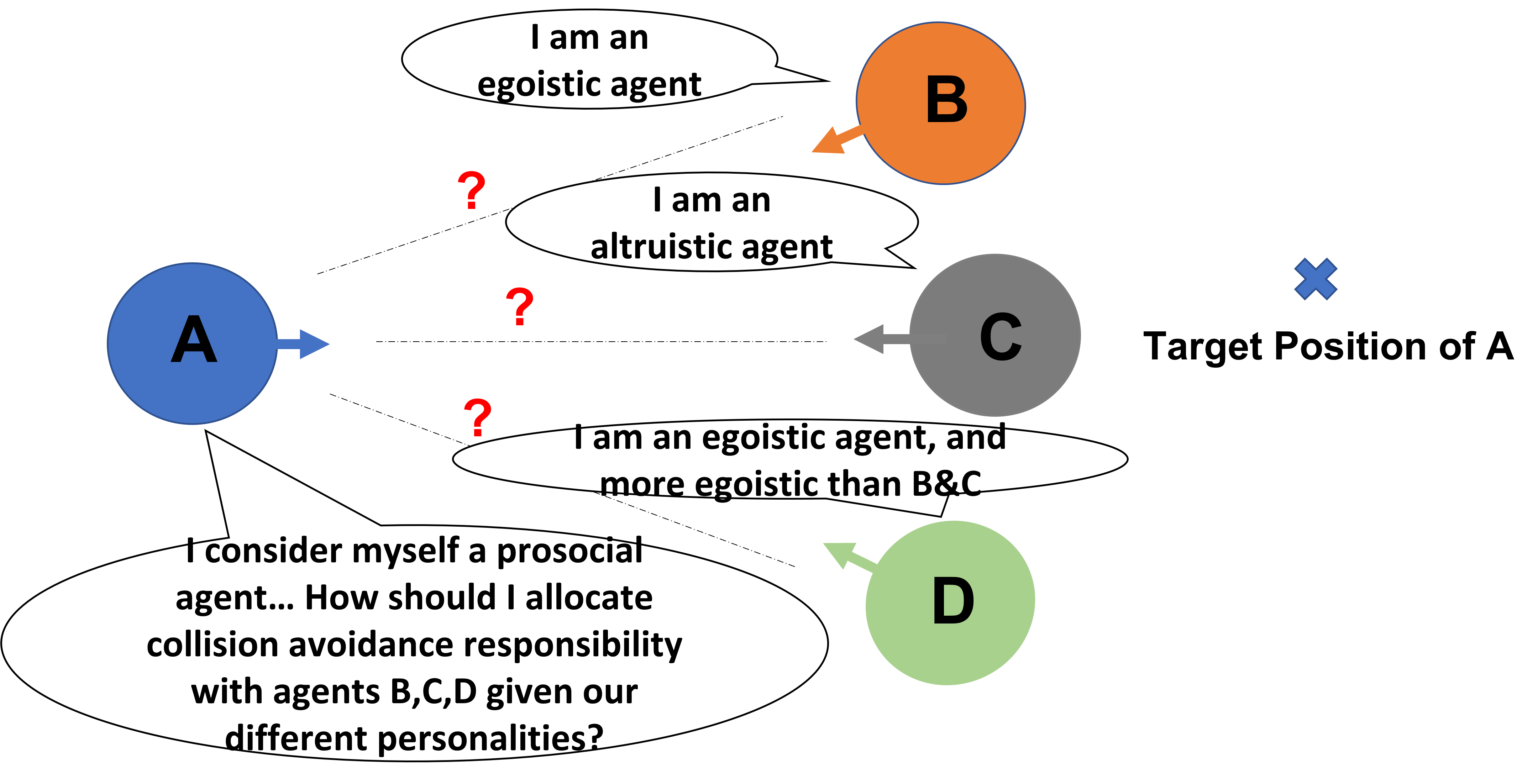}
    \caption{\footnotesize
    \label{scenario_illustration}
   A multi-agent semi-cooperative task where all agents share different social preferences or personalities. How can they achieve efficient and safe interaction?  }
\end{figure}

Multi-agent systems have been extensively studied in recent years given its potential in modelling and handling complex tasks through cooperative behaviors among agents, e.g. search and rescue \cite{parker2016exploiting}, environmental exploration \cite{rouvcek2019darpa}, and coordinated autonomous driving \cite{yu2019distributed, lyu2021probabilistic}.
One critical challenge for the success of multi-agent interaction often lies in the safety assurance when producing the inter-agent behaviors. While this has been studied in terms of safe control and collision avoidance for autonomous systems such as multi-robot systems \cite{van2008reciprocal, alonso2012reciprocal, wang2019game, luo2020multi, lyu2021probabilistic}, it is often assumed that the interacting agents in a collision avoidance scenario such as that shown in Fig. \ref{scenario_illustration}, are either fully cooperative in a symmetric manner or non-cooperative as passive moving obstacles \cite{Ames2014,luo2020multi, ames2019control,wang2017safety,lyu2021probabilistic}. Considering the possible varying coordination levels among agents in a real-world scenario, e.g. multi-vehicle traffic with egoistic and altruistic drivers showing various social preferences in participating in the collision avoidance,
it is desired to (a) model such heterogeneity in the agents' collision avoidance behaviors and (b) design the decentralized safe control framework accordingly for autonomous agents to accommodate the possible heterogeneity  and ensure safety assurance during interaction.

Collision avoidance for distributed multi-agent systems often relies on individual robot controller to constrain their own motion based on the assumed behavior models of the surrounding robots and environments. For example, reactive collision avoidance methods such as reciprocal velocity obstacles \cite{bareiss2015generalized, best2016real}, safety barrier certificates \cite{borrmann2015control, wang2017safety}, and Buffered Voronoi Cells \cite{zhou2017fast} have been extensively used to model the collision-free multi-robot behaviors by characterizing the admissible control space or local collision-free position space for each robot. 
The pre-computed safety guarantee is achieved with the assumption of fully cooperative neighboring robots under the same control design or passive obstacles moving with piece-wise constant velocity known to each robot. 
Although the heterogeneity of individual agent's behavior can be described by varying capability-related parameters such as safety radius, actuation limits, etc. As discussed in \cite{wang2017safe}, it is still challenging to model heterogeneous behavior patterns such as varying social preferences/personalities that generalize different levels of conservativeness and coordination among interacting robots in collision avoidance scenario. 


To incorporate the social preferences or personalities information into a multi-agent control framework,
some researchers draw ideas from sociology and psychology, for instance, the concept of Social Value Orientation (SVO)~\cite{Liebrand1984TheEO,murphy2011measuring} as quantified selfishness to characterize egoistic and altruistic agents. These are promising potential solutions, as concepts like SVO provide a quantitative way to measure the agent choice over weighting its own rewards against the others in social dilemmas. There has been growing interest in leveraging these concepts to design heterogeneous multi-agent interaction mechanisms, e.g. \cite{schwarting2019social, pierson2020weighted, toghi2021social}.

Motivated by these considerations, we focus on safe interaction in heterogeneous multi-agent systems. In order to model the influence of individual social preference or personality on the interactive behaviors, e.g. collision avoidance between pairwise agents, we propose  Responsibility-associated Social Value Orientation (R-SVO) as a novel variant of traditional SVO. This enables a unified representation to describe the relative social implications among agents during interaction. For example, when operating in close proximity, how likely are agents with various absolute individual selfishness measures to yield to each other for collision avoidance? With R-SVO defining such relative responsibility allocation among agents, we then propose a Control Barrier Function (CBF)-based computational framework for decentralized safe control that encodes R-SVO in terms of shared responsibility into the collision-free behavior design. Such a framework is able to accommodate the various quantified social preferences of individual agents by allocating responsibility shares accordingly when enforcing reciprocal safe constraints. The framework is proved to be valid in ensuring overall safety for the multi-agent system.

Our \textbf{main contributions} are: \textbf{1)} Drawing upon the idea of Social Value Orientation (SVO), we propose a novel concept of Responsibility-associated Social Value Orientation (R-SVO) to redefine the relative difference in social preference or personality among agents while preserving the intended social implications and other properties inherited from SVO; \textbf{2)} we use Local Pairwise Responsibility Weights to encode agents' social preference in responsibility allocation in multi-agent collision avoidance; and \textbf{3)} we present a decentralized safe control framework for individual agents to ensure provably correct collision-free interaction in multi-agent systems. Simulations on multi-agent position-swapping games are provided to demonstrate the advantages of the proposed framework in terms of saving overall task completion time and deadlock resolution efficiency.

\section{Related Work}

\begin{figure}
    \centering
    \includegraphics[width = 0.4\linewidth]{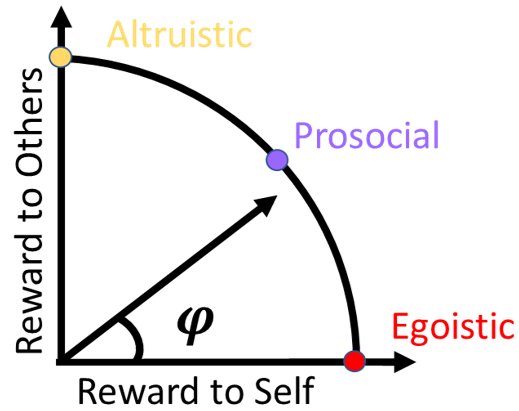}
    \caption{\footnotesize
    \label{svo}
   Concept of Social Value Orientation in angular representation \cite{schwarting2019social}. The angular value ranges from zero to $\frac{\pi}{2}$, representing the changes in weight of how an agent value its own reward against the rewards of others.}
\end{figure}

Safe control in terms of collision avoidance is critical for multi-agent systems. To minimize the deviation from robots' primary task execution due to safety considerations, reactive collision avoidance methods such as reciprocal velocity obstacles \cite{van2008reciprocal, alonso2012reciprocal},
safety barrier certificates \cite{wang2017safety, wang2017safe, luo2020multi,wang2016safety} and buffered Voronoi cells \cite{wang2019game, angeris2019fast} have been presented to minimally revise the robot's task-related controller subject to collision avoidance constraints. 
Among these methods, Control Barrier Function (CBF) \cite{ames2019control} based approaches such as safety barrier certificates \cite{wang2017safety, luo2020multi} become increasingly popular due to the advantages of rendering larger admissible control space and addressing nonlinear agent dynamics with formally provable safety guarantees. In multi-robot collision avoidance, the safety barrier certificates define an admissible safety region around individual robot's controller from which the resulting multi-robot behaviors are collision-free. To consider collision avoidance with heterogeneous agents, \cite{wang2017safe} proposed a distributed safe control framework using CBF to address robots with varying dynamics in terms of safety radius, acceleration limits, etc, yet it is difficult to design individual robot controller reflecting the varying degrees of coordination in multi-robot interaction, e.g. how to determine the proper control space of an altruistic agent when interacting with an egoistic agent for safety guarantee.



To explore interaction among heterogeneous agents with different social preferences or personalities, 
some works borrow ideas from social psychology like Social Value Orientation (SVO)~\cite{Liebrand1984TheEO,murphy2011measuring,murphy2015social} to quantify the effect of agent personalities on behavior modeling. SVO is usually represented in angular notation $\phi$ with different values corresponding to various personalities, as shown in Fig. \ref{svo}. The angular value $\phi$ ranges from zero to $\frac{\pi}{2}$, corresponding to different personalities ranging from egoistic ($\phi \in [0,\frac{\pi}{4}]$) to prosocial ($\phi =\frac{\pi}{4}$) and then to altruistic ($\phi \in (\frac{\pi}{4},\frac{\pi}{2}]$). \cite{pierson2020weighted} presents a weighted buffered Voronoi Cell method which incorporates SVO as a linear weight into the definition of the Voronoi tesselation. This enables egoistic agents to have a larger action space than less egoistic agents to avoid collisions. However, instead of analyzing how to decide the appropriate SVO value for different agents, SVO is simply assigned to agents directly for use without any justification. A more principled way to construct the appropriate SVO for heterogeneous agents in multi-agent systems is needed. \cite{schwarting2019social} estimates the SVO values of human drivers online as the coefficients in a human decision model to predict their future trajectories. However, the SVO values estimated from human drivers are only used as a way to express human intentions to the autonomous vehicle. In this paper, we introduce a novel concept called Responsibility-associated SVO (R-SVO) to measure the relative degree of personality differences among agents, and demonstrate how it can be incorporated into a CBF-based safe controller for semi-cooperative multi-agent interaction.
To the best of our knowledge, this is the first paper incorporating agent social preferences in CBF-based safe control frameworks and leveraging the measurement of relative personality difference among agents to contribute towards the responsibility allocation in multi-agent collision avoidance.

The remainder of the paper is organized as follows: Section \ref{background} reviews the background on Control Barrier Functions. Section \ref{method} presents the proposed Responsibility-associated Control Barrier Function-based decentralized control framework. Simulations and discussion of the proposed algorithm in three examples are provided in Section \ref{experiment}.


\section{Background on Control Barrier Functions}
\label{background}
Control Barrier Functions (CBF) \cite{ames2019control} is a method used to define an admissible control space for safety assurance of dynamical systems.
One of its important properties is its forward-invariance guarantee of a desired safety set. Consider the following nonlinear system in control affine form:
\begin{equation}\label{eq:nonlinear}
    \dot x = f(x)+g(x)u
\end{equation}
where $x\in \mathcal{X}\subset \mathbb{R}^n$ and $u\in\mathcal{U}\subset \mathbb{R}^m$ are the system state and control input with $f$ and $g$ assumed to be locally Lipschitz continuous.
A desired safety set $x\in\mathcal{H}$ can be denoted by a safety function $h(x)$: 
\begin{equation}\label{eq:safeset_general}
\mathcal{H} =\{x \in \mathbb{R}^n : h(x)\geq 0\}
\end{equation}
Thus the control barrier function for the system to remain in the safety set can be defined as follows \cite{ames2019control}:
\begin{definition}
(Control Barrier Function) Given a dynamical system (\ref{eq:nonlinear}) and the set $\mathcal{H}$ defined in (\ref{eq:safeset_general}) with a continuously differentiable function $h:\mathbb{R}^n\rightarrow \mathbb{R}$, then $h$ is a control barrier function (CBF) if there exists a class $\mathcal{K}$ function for all $x\in \mathcal{X}$ such that 
\begin{equation}\label{eq:cbf_def}
    \sup_{u\in\mathcal{U}} \ \{L_f h(x)+L_g h(x) u\}\geq -\kappa \big(h(x)\big)
\end{equation}
\end{definition}
\noindent
where $\dot{h}(x,u)=L_f h(x)+L_g h(x) u$ with $L_f h, L_g h$ as the Lie derivatives of $h$ along the vector fields $f$ and $g$.

A commonly selected class $\mathcal{K}$ function is $\kappa (h(x))=\gamma h(x)$~\cite{ames2019control,zeng2021safety,he2021rulebased}, where $\gamma\in\mathbb{R}^{\geq 0}$ is a CBF design parameter controlling system behaviors near the boundary of $h(x)=0$. Hence, the admissible control space in (\ref{eq:cbf_def}) can be redefined as 
\begin{equation}\label{eq:cbf}
    \mathcal{B}(x)=\{u\in\mathcal{U}:\dot{h}(x,u) + \gamma h(x)\geq 0\; \}
\end{equation}
It is proved in \cite{ames2019control} that any controller $u\in\mathcal{B}(x)$ will render the safe state set $\mathcal{H}$ forward-invariant, i.e., if the system (\ref{eq:nonlinear}) starts inside the set $\mathcal{H}$ with $x(t=0)\in \mathcal{H}$, then it implies $x(t)\in\mathcal{H}$ for all $t>0$ under controller $u\in\mathcal{B}(x)$.

\section{Method}
\label{method}

\begin{figure}
    \centering
    \includegraphics[width = 1\linewidth]{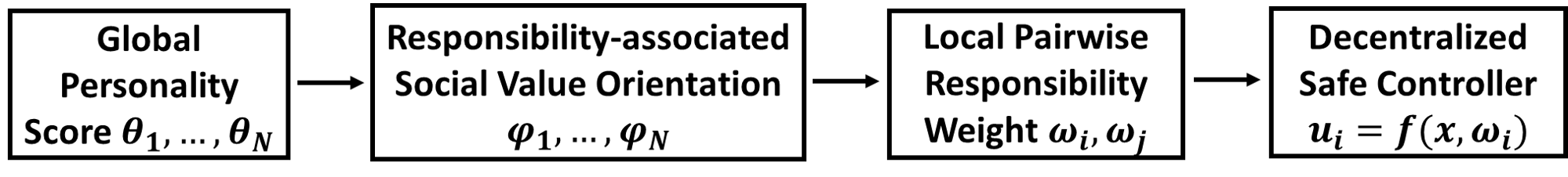}
    \caption{\footnotesize
    \label{framework}
   Proposed framework. We propose the concept of Responsibility-associated Social Value Orientation (R-SVO) to measure the relative difference in agent personalities. Local Pairwise Responsibility Weight uses the information from (R-SVO) and is applied as a weight in the CBF-based decentralized safe controller to solve the responsibility allocation problem in multi-agent collision avoidance.}
\end{figure}

In this section, we start by introducing the definition and meaning of Responsibility-associated Social Value Orientation (R-SVO), and its connection with tradition SVO in terms of properties and social implications. This is the first step to show how to construct the approriate R-SVO for agents considering the relative difference in personalities between themselves and their pairwise agents. Then we present the proposed Responsibility-associated CBF-based safe control framework, where Local Pairwise Responsibility Weight is defined to connect from R-SVO to responsibility allocation weights used in the control framework. Then detailed proofs are provided showing that the proposed framework can always formally guarantee interaction safety among all agents. The pipeline of the proposed framework is described in Fig. \ref{framework}.

\subsection{Social Value Orientation-based Responsibility Allocation}

Consider a multi-agent system with a total number of agents $N \in \mathcal{N}$. $\theta_i \in \mathbb{R}^{\geq 0}$ for $i = \{1,...,N\}$ is the global personality score of the agent $i$, representing the degree of agent egoism. The smaller $\theta_i$ is, more egoistically the agent $i$ behaves, and similarly, the larger $\theta_i$ is, more altruistic the agent $i$ is.

\textbf{Problem Statement}: \textit{How to ensure the aforementioned multi-agent system is collision-free, while taking agent personalities into consideration?}

While addressing collision avoidance within a multi-agent system, pairwise agent safety is considered. To this end, for any pair of agents $i$ and $j$, it is important to evaluate the appropriate responsibility given their relative degree of personalities. The motivation is that, although the global personality scores $\theta_i$ and $\theta_j$ are available, suppose both of them share the same large value, meaning both of them are altruistic agents, it doesn't add much information when we consider how to allocation responsibilities between agents in collision avoidance.


In this work, inspired by SVO, we propose to use the Responsibility-associated Social Value Orientation (R-SVO) to define the relative difference in personalities between pairwise agents.

\begin{definition}
For any pair of agents $i$ and $j$, given their global personality scores $\theta_i$ and $\theta_j$, the \textbf{ Responsibility-associated Social Value Orientation} (R-SVO) of each agent $\phi$ is defined as:
\begin{equation}
\label{rsvo}
\begin{split}
      \phi_i = \frac{\theta_i}{\theta_i+\theta_j}\cdot \frac{\pi}{2} \quad
    s.t. \quad \theta_i+\theta_j >0  
\end{split}
\end{equation}
\end{definition}
$\phi_i$ measures how the agent $i$ weights its reward against the reward of agent $j$.

Here, in order to verify that R-SVO $\phi_i$ is a legitimate representation of SVO, we prove 1) $\phi_i \in [0,\frac{\pi}{2}]$, and 2) the more egoistic the agent $i$ is, the smaller $\phi_i$ is.   

\begin{proof}

\textbf{1) To show R-SVO shares the same bounds of SVO, $\phi_i \in [0,\frac{\pi}{2}]$:}
$\theta_i, \theta_j \in \mathbb{R}^{\geq 0}$ indicates that $\phi_i \in \mathbb{R}^{\geq 0}$. The lower bound of $\phi_i=0$ is achieved when $\theta_i=0$. The upper bound of $\phi_i=1$ is achieved when $\theta_j=0$. 

For any $\theta_i, \theta_j \in \mathbb{R}^{>0}$,
\begin{equation}
\begin{split}
       \phi_i = \frac{\theta_i}{\theta_i+\theta_j}\cdot \frac{\pi}{2} = \frac{1}{1+\frac{\theta_j}{\theta_i}}\cdot \frac{\pi}{2} \\
       1+\frac{\theta_j}{\theta_i} \in [1,+\infty) \rightarrow \phi_i \in (0,\frac{\pi}{2}]
\end{split}
\end{equation}
Therefore, we conclude the proof that  $\phi_i \in [0,\frac{\pi}{2}]$.

 \textbf{2) The definition of R-SVO inherits the same social implication from SVO:} The smaller $\theta_i$ is, the larger $1+\frac{\theta_j}{\theta_i}$ is, the smaller $\phi_i$ is, meaning the more egoistically agent $i$ behaves. If $\theta_i=\theta_j$, 
 \begin{equation}
     \phi_i = \frac{\theta_i}{\theta_i+\theta_j}\cdot \frac{\pi}{2} =  \frac{\pi}{4}
 \end{equation}
~\cite{Liebrand1984TheEO,murphy2011measuring} also verify that $\phi = \frac{\pi}{4}$ represents the prosocial personality, which is exactly the intended social implication to express.
\end{proof}

\subsection{Responsibility-associated Control Barrier Function-based Safe Control}


In multi-agent systems, agents are usually modeled using a single integrator: $\dot{x} = u$, where $x\in \mathbb{R}^2$ is the position of the agents, and $u \in \mathbb{R}^2$ is the velocity as control input. For any pair of agents $i$ and $j$ employing Control Barrier Function-based safe controllers, we consider the particular choice of pairwise agent safety function $h(x_i,x_j)$ and safety set $\mathcal{H}$ as follows.
\begin{equation}\label{eq:multi_safety}
\begin{split}
    \mathcal{H}(x)&=\{x\in\mathcal{X}: \; h(x_i,x_j) = ||x_i-x_j||^2-R_{safe}^2\geq 0,\forall i\neq j\} 
\end{split}
\end{equation}
where $x_i,x_j \in \mathbb{R}^2$ are the positions of agents $i$ and $j$, and $R_{safe}\in \mathbb{R}$ represents the predefined safety margin. The admissible control space in Eq. \ref{eq:cbf} then becomes:
\begin{equation}
        \mathcal{B}(x)=\{u_i,u_j\in\mathcal{U}:\dot{h}(x_i,x_j,u_i,u_j) + \gamma h(x_i,x_j)\geq 0\; \}
\end{equation}
Given the agent dynamics equation, we have 
\begin{equation}
\label{eq:centralized-cbf}
    \begin{split}
        2(x_i-x_j)&^T(u_i-u_j)+\gamma h(x_i,x_j)\geq 0\\
        -2(x_i-x_j)&^Tu_i+2(x_i-x_j)^Tu_j \leq \gamma h(x_i,x_j)
    \end{split}
\end{equation}
meaning that as long as $u_i$ and $u_j$ satisfy Eq. \ref{eq:centralized-cbf}, then safety between the agent pair $i$ and $j$ is guaranteed.

Incorporating the Responsibility-associated SVO introduced in the previous section, we propose a decentralized safe control framework with responsibility allocation:
\begin{theorem}
In a multi-agent system, agent safety during an interaction is formally guaranteed, if for any pair of agents $i$ and $j$, agent $i$ takes the \textbf{Local Pairwise Responsibility Weight} $\omega_i = cos^2(\phi_i)$, 
so that the admissible control space in a centralized system in Eq. \ref{eq:centralized-cbf} is converted to: 
\begin{equation}
    \begin{split}
    \mathcal{B}(x)=&\{u_i\in\mathcal{U}_i: A_i u_i\leq \omega_i b_i, \\
    A_i = -2(x_i-x_j)^T& \in \mathbb{R}^{1\times2}, b_i = \gamma h(x_i,x_j) \in \mathbb{R}\; \}
    \end{split}
\end{equation}
\label{respobsibility-theorem}
\end{theorem}

Therefore, for any pair of agents in multi-agent interaction, the \textbf{Responsibility-associated CBF-based decentralized safe control framework} is formulated as a quadratic program:
\begin{equation}
\begin{split}
      \min_{u_i\in\mathcal{U}_i} || u_i&-\Bar{u}_i||^2 \\
    s.t \quad u_{min} \leq &u_i \leq u_{max}\\
    A_i u_i\leq \omega_i b_i, \quad
    A_i = -2(x_i&-x_j)^T, \quad    b_i = \gamma h(x_i,x_j)
\end{split}
\label{qp}
\end{equation}
where $\Bar{u_i} \in \mathbb{R}^2$ is the nominal controller input, assumed to be computed by a higher-level task-related planner, for example, a behavior planner. $u_{min}$ and $u_{max}$ are the minimum and maximum allowed velocity.

\subsection{Evaluation of Multi-agent System Safety}
In this section, detailed proof is provided to show that the proposed framework in Th. \ref{respobsibility-theorem} is the necessary condition of formally provable safety guarantees. In other words, we aim to show that for any pair of agents $i$ and $j$, Th. \ref{respobsibility-theorem} ensures the agents will not collide.

\begin{proof}

\textbf{Step 1:} By applying Th. \ref{respobsibility-theorem} to agent $j$, we get the safety constraint over $u_j$ as:
\begin{equation}
    A_ju_j\leq\omega_j b_j, \quad A_j = -2(x_j-x_i)^T, \quad b_j = \gamma h(x_j,x_i)
    \label{agentj-responsibility}
\end{equation}
We know $A_j = -2(x_j-x_i)^T = 2(x_i-x_j)^T$. The summation of the left hand sides of the inequalities in Th. \ref{respobsibility-theorem} and Eq. \ref{agentj-responsibility} is:
\begin{equation}
    \begin{split}
        A_i u_i + A_ju_j 
        = -2(x_i-x_j)^T u_i+2(x_i-x_j)^T u_j
    \end{split}
\end{equation}
which is exactly the left hand side of the inequality in Eq. \ref{eq:centralized-cbf}.

\textbf{Step 2:} By Eq. \ref{eq:multi_safety}, we have $b_j = h(x_j,x_i) = h(x_i,x_j) = b_i$, and now we sum the right hand sides of the two inequalities and get:
\begin{equation}
    \omega_i b_i+ \omega_j b_j = (\omega_i+\omega_j) b_i = (\cos^2(\phi_i)+\cos^2(\phi_j))\gamma h(x_i,x_j)
\end{equation}
As long as we can prove that $(\cos^2(\phi_i)+\cos^2(\phi_j))\gamma h(x_i,x_j)$ is an equal or a tighter bound compared to $\gamma h(x_i,x_j)$, which is the right hand side of Eq. \ref{eq:centralized-cbf}, safety between pairwise agents during interaction can be formally guaranteed.

\textbf{Step 3:} By Def. \ref{rsvo}, we have $\phi_i = \frac{\theta_i}{\theta_i+\theta_j}\cdot\frac{\pi}{2}$ and $\phi_j = \frac{\theta_j}{\theta_i+\theta_j}\cdot\frac{\pi}{2}$, so $\phi_j = \frac{\pi}{2}-\phi_i$. Since $\phi_i \in [0,\frac{\pi}{2}]$,
\begin{equation}
    \begin{split}
        \cos^2(\phi_j) &= \cos^2(\frac{\pi}{2}-\phi_i) = \sin^2(\phi_i)\\
\Rightarrow \qquad \qquad  (\cos^2&(\phi_i)+\cos^2(\phi_j))\gamma h(x_i,x_j) \\
        = (\cos^2(\phi_i)+&\sin^2(\phi_i))\gamma h(x_i,x_j)=\gamma h(x_i,x_j)
    \end{split}
\end{equation}
Thus the proof is concluded that the proposed Responsibility-associated decentralized CBF in Th. \ref{respobsibility-theorem} provides an equivalent safety guarantee to that of the common centralized CBF in Eq. \ref{eq:centralized-cbf}.
\end{proof}




The algorithm of the proposed framework is presented in Algorithm~\ref{alg:responsibility_ctrl}.
\begin{algorithm}
\caption{Responsibility-associated CBF-based \\Decentralized Safe Control Framework}
\begin{algorithmic}\label{alg:responsibility_ctrl}
\REQUIRE $\theta_{1,...,N}, x^0_{1,...,N}, \bar{u}^t_{1,...,N}, \gamma, R_{safe}$
\ENSURE $u^t_{1,...,N}$
\FOR{$t = 1:T$}
\FOR{$i=1:N$}
\FOR{$j=1:N$ except $i$}
\STATE
Compute Responsibility-associated Social Value Orientation $\phi_i$ (Eq. \ref{rsvo})
\STATE
Compute Local Pairwise Responsibility
Weight $\omega_i$ (Th. \ref{respobsibility-theorem})
\STATE Calculate pairwise agent safety function $h(x_i,x_j)$ Eq. \ref{eq:multi_safety}
\STATE Calculate $A_i^t,b_i^t$ for safety constraint composition (Eq. \ref{qp})
\STATE Stack $A^t_i,b^t_i$ for all surrounding agent $j$
\ENDFOR
\STATE Minimum Deviation Control:  $\min_{u^t_i} || u^t_i-\Bar{u}^t_i||^2$ \quad
\STATE
    s.t $\quad u^t_i \in [u_{min}, u_{max}],\quad A^t_i u^t_i\leq \omega_i b^t_i$
\ENDFOR
\ENDFOR
\end{algorithmic}
\end{algorithm}
The superscript $0$ represents the initial condition at $t = 0$, and the superscript $t$ represents variables at timestep $t$. At each timestep, we consider all pairwise agents for safety constraint composition. For every agent $i$ with all its pairwise agents $j$, R-SVO $\phi_i$ is calculated based on the global personality score $\theta_i$ and $\theta_j$ to measure the personality difference between agents. Then Local Pairwise Responsibility Weight $\omega_i$ is computed based on R-SVO $\phi_i$. Using the Local Pairwise Responsibility Weight, responsibility allocation is achieved between the pairwise agents in collision avoidance, and the decentralized safety-constraint over $u_i^t$ is composed with state (position) observations of both agents $x_i^t$ and $x_j^t$. The proposed framework scales up well with a larger number of agents and is highly generally applicable to other real-time robotics applications.

\section{Simulation \& Discussion}
\label{experiment}
In this section, simulations using Matlab are provided to demonstrate the validity and effectiveness of the proposed method. Three provided examples include: a two-agent position swapping game, a self-driving car highway ramp merging problem, and a multi-agent circular position swapping game.

\subsection{Pairwise-agent Interaction: Position Swap Game}
\begin{figure}
    \centering
    \includegraphics[width = 1\linewidth]{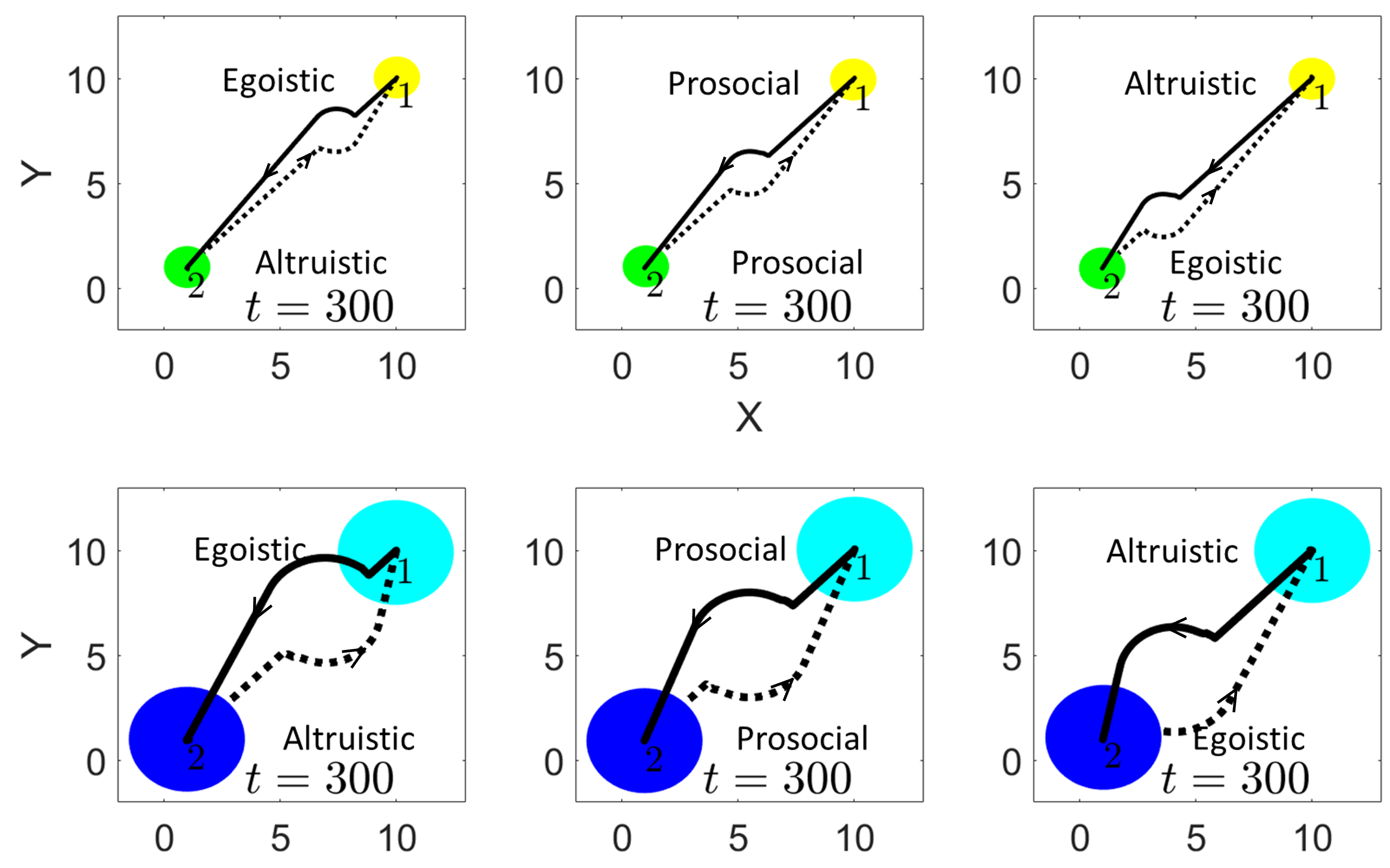}
    \caption{\footnotesize
    \label{2-agent-exp}
  Final positions of two agents in a position Swapping Game. The dashed line represents the trajectory of agent $1$, which starts from the lower left corner, and the solid line represents the trajectory of agent $2$, which starts from the upper right corner. The colorful circles stands for predefined safety margin for agents, with  $R_{safe} = 1$ for the first row and $R_{safe} = 2$ for the second row. The three columns corresponds to three different Local Pairwise Responsibility Weight settings.}
\end{figure}

A two-agent  position swapping game is demonstrated to show how agent personalities contribute to the pairwise agent responsibility allocation.
Both agents aim to navigate to their goal locations without colliding with each other while employing a move-to-goal controller $\Bar{u} = -k\cdot(x-x_{target})$, where $k\in \mathbb{R}^{>0}$, and $x_{target}\in\mathbb{R}^2$ is the goal position of each agent.
A right hand heuristic rule is designed to avoid deadlock situations as in \cite{pierson2020weighted}. The final result is shown in Fig. \ref{2-agent-exp}. Two sets of trajectories of pairwise agents with different predefined safety margins $R_{safe}=1, R_{safe}=2$ are shown in two rows respectively, where the three columns represent three kinds of responsibility-associated behavior corresponding to agents' various personalities. The dashed line represents the trajectory of agent $1$, which starts from the lower left corner, and the solid line represents the trajectory of agent $2$, which starts from the upper right corner. 

In the first set of experiments (first column), agent $1$ is an egoistic agent compared to agent $2$ ($\theta_1<\theta_2$), and the resulting Local Pairwise Responsibility Weights are $\omega_1 = 0.8, \omega_2 = 0.2$, which also means that $u_2$ is enforced by a tighter bound than $u_1$ when considering collision avoidance. Therefore, the two agents first move straight towards each others' initial locations, with agent $1$ moving at a faster velocity than agent $2$, since it enjoys a relatively flexible bound over $u_1$. Once the two agents come to a point where deadlock is about to happen, by applying the right hand heuristic rule, they take appropriate actions to avoid collision.

In the second set of experiments (second column), agent $1$ shares the same global personality as agent $2$ ($\theta_1=\theta_2$), leading to Local Pairwise Responsibility Weights $\omega_1 = 0.5, \omega_2 = 0.5$, so that both agents share the same portion of responsibility, leading to symmetric behavior. In the third set of experiments (third column), agent $1$ is an altruistic agent compared to agent $2$ ($\theta_1>\theta_2$), thus the Local Pairwise Responsibility Weights are $\omega_1 = 0.2, \omega_2 = 0.8$. Benefiting from a more flexible control constraint, Agent $2$ moves at a higher velocity and proceeds a longer distance than agent $1$ before they meet. Generally, the responsibility allocation between two agents is easier to observe in examples with larger safety margin $R_{safe}$. The above examples show how the proposed method generates various responsibility allocations in producing safe interaction behaviors based on different agent personalities.

\subsection{Examples in Highway Ramp Merging Scenario}
\begin{figure}
    \centering
    \includegraphics[width = 1\linewidth]{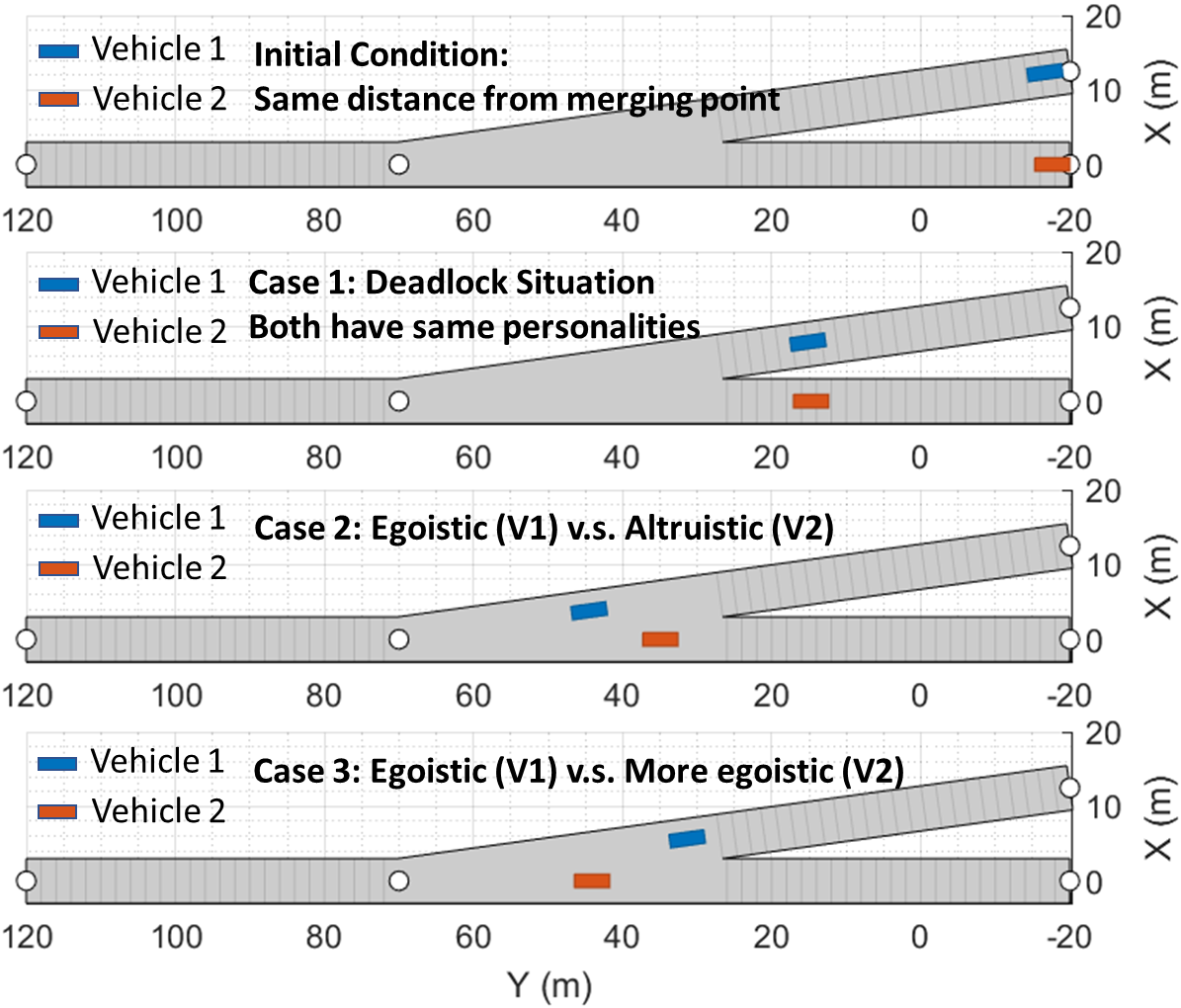}
    \caption{\footnotesize
    \label{av-example}
   Application of the proposed method in the highway ramp merging scenario.}
\end{figure}

The proposed framework is generally applicable to different robotics applications which may require coordination with agent social preferences or personalities, such as autonomous driving, collaborative manufacturing robots, and other general mobile robots. 
In Fig. \ref{av-example}, we show a ramp merging scenario where two vehicles start from their initial location on each lane with the same distance away from the fixed ramp merging point at $Y = 70m$. The goal for each vehicle is to pass the merging point as soon as possible without any collision. 
Three cases are given as examples to demonstrate how the resulting safe behaviors of the vehicles $V1,V2$ using our proposed decentralized method align with vehicle's individual personality/social preference (e.g. egoistic, altruistic, etc) during interaction. 

The system dynamics of a vehicle is described by double integrators as in \cite{lyu2021probabilistic}: 
\begin{equation} 
\begin{split}
    \dot{X} &=\begin{bmatrix}
    \dot{x}\\
    \dot{v}
    \end{bmatrix}
    =\begin{bmatrix}
    0_{2\times2}\; I_{2\times2}\\
    0_{2\times 2} \;0_{2\times 2}
    \end{bmatrix}
    \begin{bmatrix}
    x \\
    v
    \end{bmatrix}
    + \begin{bmatrix}
    0_{2\times2}\; 0_{2\times2}\\
    I_{2\times2}\; 0_{2\times2}
    \end{bmatrix}\begin{bmatrix}
    u\\ 0
    \end{bmatrix} \\
\end{split}
\label{dynamics}
\end{equation}
where $x\in\mathbb{R}^2,v\in \mathbb{R}^2$ are respectively the position and linear velocity of each vehicle. $u\in\mathbb{R}^2$ represents the acceleration control input that is determined by our proposed decentralized responsibility-associated safe controller (Algorithm~\ref{alg:responsibility_ctrl}).

In case 1 when two vehicles $V1,V2$ have the same personality/social preference with the corresponding global personality scores $\theta_{1}=\theta_{2}$, our Algorithm~\ref{alg:responsibility_ctrl} assigns the same Local Pairwise Responsibility Weights value $\omega_{1}=\omega_2=0.5$ for computing the decentralized vehicle controller. Hence it degenerates to the homogeneous safe controller as in Fig.~\ref{2-agent-exp} (middle column) and as expected, deadlock situation happens in this symmetric setting, no matter whether $V1$ and $V2$ are both egoistic vehicles or altruistic vehicles.

Similar to the example in Fig.~\ref{2-agent-exp} (left column), in case 2 in Fig.~\ref{av-example} when $V2$ is altruistic compared to $V1$, 
$V1$ then has a less restrictive motion in avoiding collisions due to the assigned smaller R-SVO by Algorithm~\ref{alg:responsibility_ctrl}, and thus allowing it to take more aggressive actions to pass the merging point first while ensuring safety, which matches the expectation based on the relative personality between the two.
In case 3, although $V1$ is an egoistic vehicle, $V2$ is even more egoistic than $V1$. Therefore, $V2$ arrives the merging point first, while $V1$ behaves more conservatively by enforcing a tighter bound on its admissible action space.

To that end, from the three examples in Fig.~\ref{av-example} it is justified the safe behavior generated by the proposed controller is well aligned with the agent social preferences/personalities.

\subsection{Multi-agent Interaction: Circular Game}
In this section, we apply the proposed framework in a multi-agent interaction scenario, where three agent pairs, six agents in total, conduct a circular game where all agents conduct position swapping with their paired agent while ensuring no collisions with all other agents. The purpose of this example is to demonstrate how the proposed method can efficiently solve the deadlock situation in a crowded environment while remaining collision-free.

\begin{figure}
    \centering
    \includegraphics[width = 1\linewidth]{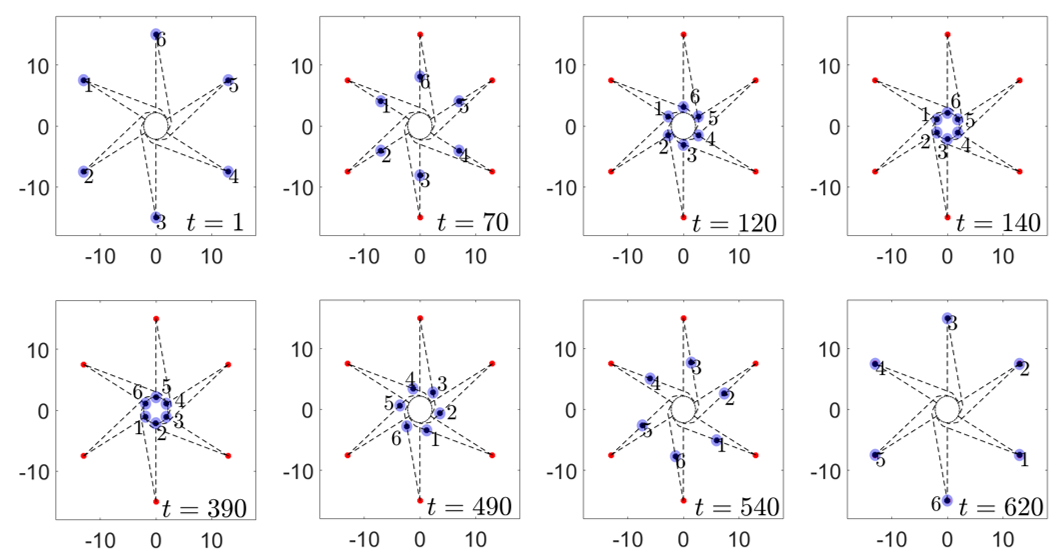}
    \caption{\footnotesize
    \label{symmetric}
   Symmetric example where no agent personality information is taken into considerations. The dash lines retrospectively show the path of the agents swapping their positions with the agents on their opposite direction, e.g. agent 1 swaps its position with agent 4. The safety margin $R_{safe}$ is set to be $1$m for all agents. The coordinate axes are in meters, and the timestep is 0.01 seconds.}
\end{figure}

In the first example, a symmetric coordination is deployed, meaning no agent personality information is taken into consideration, to serve as a comparison baseline to the proposed method. The simulation result is shown in Fig. \ref{symmetric}. It is observed that at $t = 140$, a deadlock situation is about to happen. All the six agents forms a circle, and by applying the right hand heuristic rule, all agents rotate slowly around the origin in the circular formation. The deadlock resolution ends at $t = 390$. The whole coordination task is completed at $t = 620$ and no collision happens.

\begin{figure}
    \centering
    \includegraphics[width = 1\linewidth]{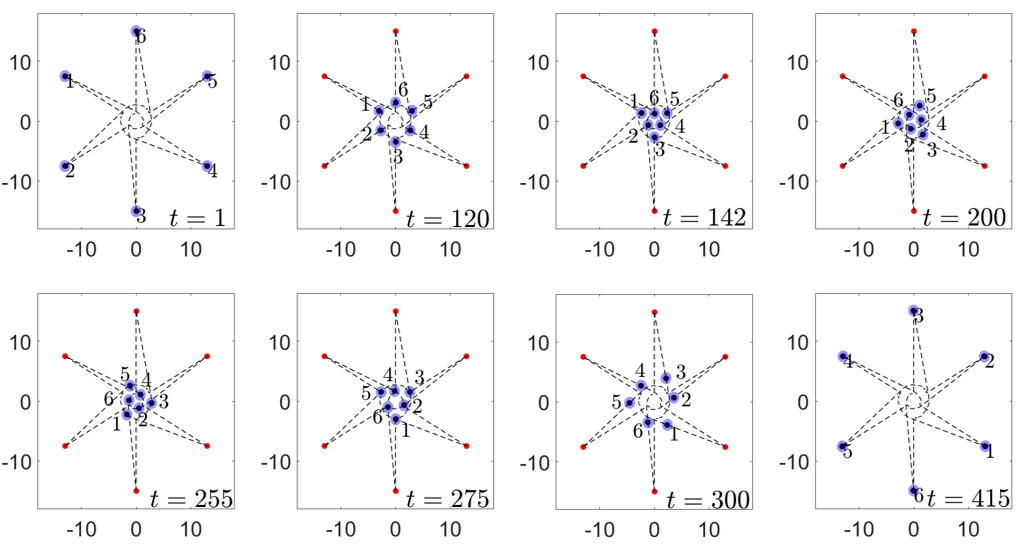}
    \caption{\footnotesize
    \label{asymmetric}
   Asymmetric example where agent $\{1,3,5\}$ are designed to be more altruistic with a larger $\theta$ and agent $\{2,4,6\}$ are designed to be more egoistic with a smaller $\theta$. The dash lines retrospectively show the path of the agents swapping their positions with the agents on their opposite direction, e.g. agent 1 swaps its position with agent 4.}
\end{figure}

    The result of the second example where the proposed asymmetric method is deployed is shown in Fig. \ref{asymmetric}. All position swapping pairs consist of an egoisitic agent and an altruistic agent. The deadlock situation occurs at $t=142$. An interesting observation is that the way agents form the deadlock resolution is quite different from the symmetric case in Fig. \ref{symmetric}. Agents $\{1,3,5\}$ form an outer triangle outside the inner triangle formed by agents $\{2,4,6\}$. During the deadlock resolution period, the relatively egoistic agents $\{2,4,6\}$ only travel half of the distance travelled by agents $\{1,3,5\}$, who are relatively altruistic. With fewer agents assigned to more than one rotation track, the rotation speed is greatly improved as congestion is eliminated compared to the previous symmetric example. The deadlock is resolved at $t=275$ and the whole circular position swapping task is completed at $t=415$ with no collision happening among agents. 
    
\begin{figure}
    \centering
    \includegraphics[width = 0.9\linewidth]{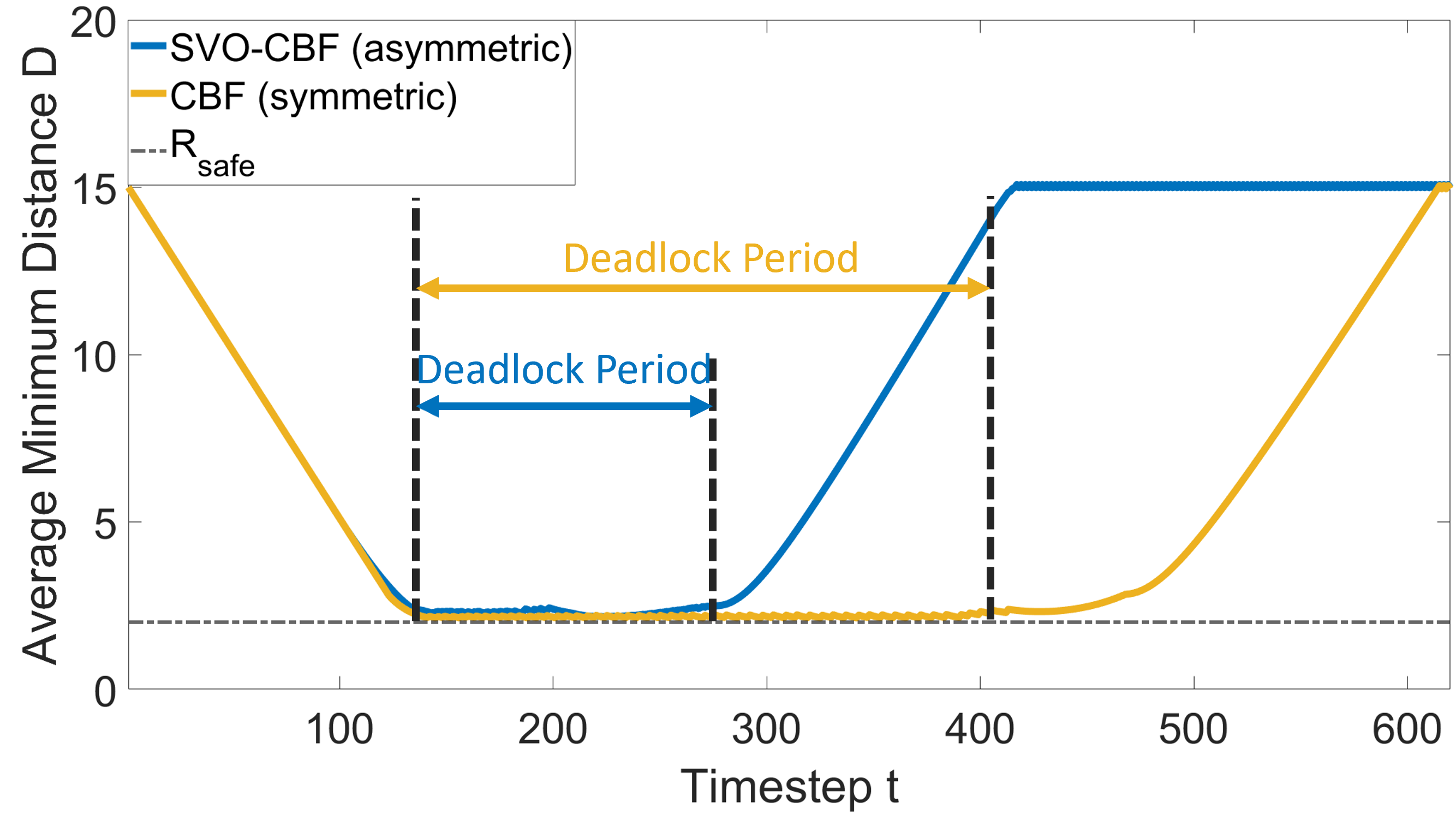}
    \caption{\footnotesize
    \label{asymmetric-distance}
   Comparison of the proposed method and symmetric control. The x-axis is timestep and y-axis is the average minimum distance among all agents in the multi-agent system.}
\end{figure}

    A more general comparison of the proposed method and symmetric control is provided in Fig. \ref{asymmetric-distance}, where the average minimum distance among agents over time is presented. Five sets of global personality scores are randomly initialized in the range of 1 to 10 excluding repeated values, and tested by applying symmetric control and the proposed method. The average minimum distance $D$ is computed for all five sets of simulation. Overall, both symmetric and asymmetric examples ensure collision-free interactions among agents. However, the proposed method outperforms the symmetric example in terms of deadlock resolution efficiency and the overall time it takes to complete the multi-agent coordination task. The deadlock resolution efficiency is improved by 46.8\% and the overall task completion time is improved by 33\%.

\subsection{Who benefits the most?}
An interesting question to ask is, who benefits the most from the multi-agent coordination task: the egoistic agents or the altruistic agents? 

We argue that both kinds of agents have their own advantages. Egoistic agents gain a relatively flexible control space when considering avoiding potential collision with others, and therefore their major tasks are less interrupted, for example, there is less deviation from the nominal controller. Altruistic agents have tighter safe control bounds in order to keep them safe, and as a result, they can deviate from the nominal controller more frequently. 

From another angle, the effect of different personalities on resulting behavior can be viewed as a kind of balance between time and travel distance. Egoistic agents stay close to the nominal control and the preferred path. However, this property may sometimes lead to higher time cost. While the altruistic agents are willing to actively accelerate deadlock resolutions, it comes with a consequence of longer travelled distance.

 


\section{CONCLUSIONS}
\label{conclusion}

In this paper, we present a novel social preference-aware decentralized safe control framework. Responsibility allocation in collision avoidance is conducted in a multi-agent system where heterogeneous agents are specified by their individual personalities. Responsibility-associated Social Value Orientation (R-SVO) is introduced to measure the relative difference in personality among agents while preserving the intended social implications and other inherited properties from traditional SVO. By leveraging the idea of proposed Local Pairwise Responsibility Weights, a Responsibility-associated Control Barrer Function-based safe control framework is presented along with formal proofs on safety guarantees. The proposed method is generally applicable to many real-time robotics applications where multi-agent coordination is required in a safe and efficient manner. In future, we aim to investigate how personalities could be assigned in the multi-agent system for better task coordination.


\bibliography{submission}{}
\bibliographystyle{IEEEtran}

\end{document}